\documentclass[legalpaper,twocolumn]{article}

\usepackage{amssymb,amsfonts,amsmath,amstext,amsgen,amsopn,amsxtra,indentfirst,graphicx}

\begin{document}

\title{Atomic iron and titanium in the atmosphere \\ of the exoplanet KELT-9b}
\date{}
\maketitle
\vspace{-0.5in}
\noindent
\author{H. Jens Hoeijmakers$^{1,2}$, David Ehrenreich$^1$, \\ Kevin Heng$^{2,7}$, Daniel Kitzmann$^2$, \\ Simon L. Grimm$^2$, Romain Allart$^1$, Russell Deitrick$^2$, Aur\'{e}lien Wyttenbach$^1$, Maria Oreshenko$^2$, \\ Lorenzo Pino$^1$, Paul B. Rimmer$^{3,4}$, \\ Emilio Molinari$^{5,6}$, Luca Di Fabrizio$^5$}\\
\author{\scriptsize \textbf{1}: Observatoire astronomique de l'Universit\'{e} de Gen\`{e}ve, 51 chemin des Maillettes, 1290 Versoix, Switzerland}\\
\author{\scriptsize \textbf{2}: University of Bern, Center for Space and Habitability, Gesellschaftsstrasse 6, CH-3012, Bern, Switzerland}\\
\author{\scriptsize \textbf{3}: University of Cambridge, Cavendish Astrophysics, JJ Thomson Ave, Cambridge CB3 0HE, United Kingdom}\\
\author{\scriptsize \textbf{4}: MRC Laboratory of Molecular Biology, Francis Crick Ave, Cambridge CB2 OQH, United Kingdom}\\
\author{\scriptsize \textbf{5}: INAF FGG, Telescopio Nazionale Galileo, Rambla Jos\'{e} Ana Fern\'{a}ndez P\'{e}rez, 7, 38712 Bre\~{n}a Baja, TF - Spain}\\
\author{\scriptsize \textbf{6}: INAF Osservatorio Astronomici di Cagliari, Via della Scienza 5 - 09047 Selargius CA, Italy}\\
\author{\scriptsize \textbf{7}: Corresponding author: kevin.heng@csh.unibe.ch}

\textbf{The chemical composition of an exoplanet is a key ingredient in constraining its formation history \cite{oberg11,madhu14,ob16}.  Iron is the most abundant transition metal, but has never been directly detected in an exoplanet due to its highly refractory nature.  KELT-9b (HD 195689b) is the archetype of the class of ultra-hot Jupiters that straddle the transition between stars and gas-giant exoplanets and serve as distinctive laboratories for studying atmospheric chemistry, because of its high equilibrium temperature of $4050 \pm 180$ K \cite{gaudi17}. These properties imply that its atmosphere is a tightly constrained chemical system that is expected to be nearly in chemical equilibrium \cite{kitzmann18} and cloud-free \cite{heng16,stevenson16}.  It was previously predicted that the spectral lines of iron will be detectable in the visible range of wavelengths \cite{kitzmann18}. At these high temperatures, iron and several other transition metals are not sequestered in molecules or cloud particles and exist solely in their atomic forms \cite{kitzmann18}. Here, we report the direct detection of atomic neutral and singly-ionized iron (Fe and Fe$^+$), and singly-ionized titanium (Ti$^+$) in KELT-9b via the cross-correlation technique \cite{snellen10} applied to high-resolution spectra obtained during the primary transit of the exoplanet.}

Motivated by the theoretical predictions \cite{kitzmann18}, we conducted a search for metal lines in the high-resolution transmission spectrum of KELT-9b, which we previously observed with the HARPS-North (HARPS-N) spectrograph during a single transit of the exoplanet. HARPS-N is a fiber-fed spectrograph, stabilised in pressure and temperature, mounted on the 3.58-meter Telescopio Nazionale Galileo (TNG) located on the Canary Island of La Palma, Spain. We recorded 19 and 30 spectra during the 3.9-hour-long transit and outside of it, respectively, covering the whole night from 31 July 2017 to 1 August 2017. Spectra were reduced with the HARPS-N data reduction software (DRS) version 3.8. They consist of 69 orders covering the wavelength range from 3874 to 6909~\AA\ and partially overlapping at their edges.  Orders were extracted individually, flat-fielded using calibrations obtained during twilight, and deblazed and wavelength-calibrated in the Solar System barycentric rest frame.  A common reference wavelength grid with a spectral resolution of 0.01 \AA\ is built. The calibrated orders are then binned on this common grid. Pixels in overlapping orders at the same wavelength are then averaged in order to conserve the flux.  Contamination of the spectra by absorption lines of the Earth's atmosphere (mainly by water) were corrected using established methods (see Methods).  The optical range of wavelengths is populated with electronic transitions of the atoms and ions of metals. With a spectral resolving power of about 115,000, these data are ideal for searching for the absorption signatures associated with metals present in the atmosphere of KELT-9b, by cross-correlating the high-resolution spectra \cite{snellen10,brogi12} with theoretical templates constructed from the cross sections of the relevant species \cite{kurucz17} (Figure \ref{fig:theory}).

We specifically searched for the spectral lines of Fe, Fe$^+$, neutral titanium (Ti) and Ti$^+$ in the data, because the relative abundances of neutral and singly-ionized iron and titanium  vary by many orders of magnitude between 2500 to 6000 K (Figure \ref{fig:theory}).  Fe$^+$ becomes more abundant than Fe at about 3900 to 4300 K, depending on whether chemical equilibrium is assumed (panel \textbf{c} of Figure \ref{fig:theory}) or if photochemistry and vertical mixing are present (panel \textbf{d} of Figure \ref{fig:theory}); Ti$^+$ becomes more abundant than Ti at about 3000 to 3400 K.  These estimates assume solar metallicity, which is the multiplicative factor that is commonly applied to the elemental abundances (C/H, O/H, N/H, Ti/H, Fe/H, etc) in our chemistry models.  This factor is unity for solar metallicity.  Furthermore, the abundances of Fe$^+$ and Ti$^+$ stabilise above these transition temperatures (Figure \ref{fig:theory}), implying that only lower bounds on the temperature may be obtained.

Having to observe through the atmosphere of the Earth means the measured transmission spectrum lacks an absolute empirical normalisation.  This lack of an absolute empirical normalisation implies that absolute atomic abundances cannot be extracted from the data, which in turn implies that the metallicity cannot be inferred from these data alone.  Additionally, model transmission spectra lack an absolute theoretical normalisation \cite{bs12,g14,hk17}.  In most hot Jupiters, the spectral continuum is dominated by a combination of Rayleigh scattering associated with molecular hydrogen, the spectral line wings of molecules and the alkali metals (which are mediated by pressure broadening), collision-induced absorption associated with helium and molecular hydrogen, and clouds or hazes.  In KELT-9b, the unusually high temperatures imply that all species are in the gas phase, and that opacity by bound-free absorption associated with hydrogen anions (H$^-$) \cite{john88} is the dominant source of the spectral continuum \cite{arc18}, over Rayleigh scattering associated with both atomic and molecular hydrogen (Figure \ref{fig:theory}).  The line wings of Fe, Fe$^+$, Ti, and Ti$^+$ are also sub-dominant (Figure \ref{fig:theory}), implying that pressure broadening is not a source of uncertainty.   

This rather peculiar property of KELT-9b, and ultra-hot Jupiters in general, allows us to meaningfully compute the relative differences between the line peaks of Fe, Fe$^+$, Ti, and Ti$^+$ and the spectral continuum provided by H$^-$.  We construct four templates using theoretical transmission spectra \cite{hk17} consisting separately of Fe, Fe$^+$, Ti, and Ti$^+$ lines, each with a H$^-$ continuum.  The ability to determine the relative normalisation between the line peaks and the continuum implies that we are able to give more weight to strong lines and less weight to weak lines in our templates, which in turn produces a higher signal-to-noise (S/N) in our cross correlations compared to an analysis using templates where all of the lines are weighted equally (i.e., a so-called ``binary mask").

Encoded in the data are two distinct signatures: the rotation of the star ($v \sin{i} = 111.4$~km~s$^{-1}$) and the orbital velocity of the exoplanet ($\pm 81$~km~s$^{-1}$), both projected along the line of sight to the observer.  Each location on the stellar disk corresponds to a different projected rotational velocity of the star.  As the exoplanet moves across the stellar disk, it traces out a range of projected rotational velocities with time (Figure \ref{fig:obs}).  This leaves an imprint on each stellar absorption line that shows up as an enhancement in flux at the wavelength corresponding to the projected rotational velocity.  When the time-averaged stellar spectrum is subtracted from the data and the cross-correlation analysis is performed, a time-dependent residual known as a ``Doppler shadow" remains (Figure \ref{fig:obs}). The tilt of the Doppler shadow with respect to the axes of time and wavelength can be used to infer the angle between the rotational axis of the star and the orbital plane of the exoplanet.  This Doppler tomography technique was previously used to infer that KELT-9b resides in a near-polar orbit \cite{gaudi17}.  Superimposed on the Doppler shadow is the time-dependent absorption spectrum of the atmosphere of KELT-9b (Figure \ref{fig:obs}).  The spectrum shifts in wavelength with time, because it is associated with the varying projected orbital velocity of the exoplanet. 

To perform the cross correlations, we subdivided the reduced spectra into 20-nm bins (for computational efficiency), each to be treated independently in the analysis.  We measured the stellar spectrum by averaging the out-of-transit exposures, divided the obtained master stellar spectrum out of each exposure and corrected for residual fluctuations in the continuum by normalizing each spectrum using a smoothing filter (with a width of 0.75 \AA).  Each spectral pixel is weighted by the reciprocal of its variance in time \cite{snellen10}.  The Doppler shadow is subtracted in order to isolate the atmospheric signal from the exoplanet (see Methods).  We cross-correlated the residuals between radial velocities of $\pm 1000$ km s$^{-1}$ with the templates.   

The cross-correlation function (CCF) peaks of Fe, Fe$^+$ and Ti$^+$ are seen as bright streaks across the axes of time and systemic radial velocity (Figure \ref{fig:detections}).  Intriguingly, these streaks are also seen in the figures of the discovery study \cite{gaudi17}, which are presumably caused by spectral lines that the exoplanet and the star have in common, notably Fe$^+$.  Co-adding the signal along these streaks, in the rest frame of the exoplanet, results in significant detections of CCF peaks for Fe, Fe$^+$, and Ti$^+$ with signal-to-noise ratios (SNRs) of 7, 14 and 9, respectively.  The CCFs are normalised by the standard deviation at velocities away from the rest frame of the exoplanet in order to produce the SNRs of the detections.  By fitting Gaussians to these CCFs, we find weighted average line contrasts of $(0.28\pm0.03)\times 10^{-3}$ ($9.3\sigma$), $(2.21\pm0.08)\times 10^{-3}$ ($26\sigma$) and $(1.28\pm0.07)\times 10^{-3}$ ($18\sigma$) for Fe, Fe$^+$ and Ti$^+$, respectively. There is no significant detection of the CCF for Ti with $(0.18\pm0.05)\times 10^{-3}$ ($3.3\sigma$).

That the lines of Fe$^+$ show up more prominently than Fe suggests that the atmospheric temperatures probed exceed $\sim 4000$ K.  The non-detection of Ti (Figure \ref{fig:detections}) supports this interpretation.  The discovery of Fe, Fe$^+$ and Ti$^+$ in KELT-9b sets the stage for the future search for carbon monoxide (CO) and water (H$_2$O) in the near-infrared range of wavelengths.  At these elevated temperatures, CO is expected to be the dominant molecule, which implies that its abundance directly mirrors the value of C/H (for $\mbox{C/O}<1$) or O/H (for $\mbox{C/O}>1$) \cite{kitzmann18}.  Detecting H$_2$O would allow one to distinguish between the carbon-poor versus carbon-rich scenarios \cite{madhu12,ht16}.  Inferring C/H and O/H allows for the metallicity to be constrained.  These prospects ensure that KELT-9b will remain an important laboratory for studying extrasolar atmospheric chemistry using both space- and ground-based telescopes.

\section*{Methods}

\begin{figure*}
\begin{center}
\includegraphics[width=2.0\columnwidth]{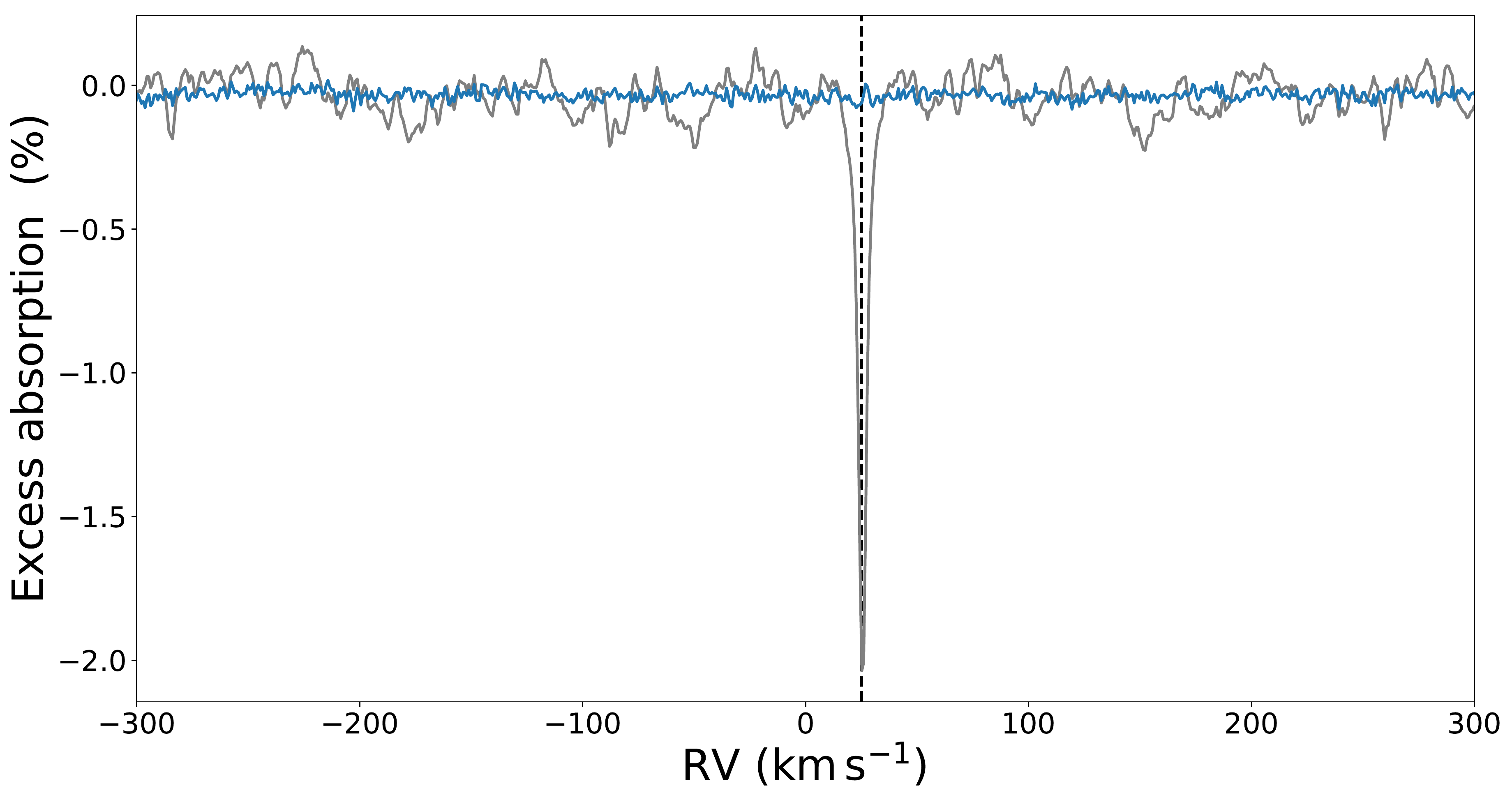}
\end{center}
\caption{Cross correlation analysis performed on the in-transit spectrum with a telluric water-absorption template spectrum (at 296 K).  The blue and grey curves are the transit depths with and without the telluric correction, respectively.  The data are shifted to the rest frame of the star, such that the signal dominated by telluric water absorption occurs near $+25$ km s$^{-1}$ (corresponding to the Barycentric Earth Radial Velocity or BERV systemic velocity), as indicated by the dashed vertical line.}
\label{fig:tellurics}
\end{figure*}

Telluric lines are the main contaminant in the observed spectrum with strong spectral lines of water and molecular oxygen in the visible range of wavelengths.  To perform decontamination, we used  version 1.5.1 of \texttt{Molecfit} \cite{smette15}, an ESO tool to correct for the tellurics in ground-based spectra following established procedures \cite{allart17}.  We tested the performance of our telluric correction by cross-correlating the corrected and uncorrected in-transit spectra with a telluric water-absorption template spectrum (at 296 K).  Figure \ref{fig:tellurics} demonstrates that our correction efficiently removes the telluric water signal.  

To remove the Doppler shadow, we cross-correlated the data with a \texttt{PHOENIX} stellar model template that matches the effective temperature of the star \cite{Husser}.  We modelled and fitted the main Doppler shadow that occurs near $-35 $~km~s$^{-1}$, as well as three aliases near $+80$, $+20$ and $-125$~km~s$^{-1}$, as time-dependent Gaussians.  The best-fit parameters (central radial velocity, width, and amplitude) are described by first-, second-, third- or fourth-degree polynomials of time.  The resulting model of the Doppler shadow is adjusted in amplitude to the CCF obtained for each species and subtracted.  We apply the cross-correlation formalism as used by \cite{allart17} to measure the average line strengths of Fe, Fe$^+$, Ti and Ti$^+$. For this purpose, the templates that were used to detect the species were continuum-subtracted to act as a weighted binary mask that measures the average spectral line present in each spectral bin and each exposure.  The final one-dimensional CCFs are obtained by a weighted co-addition of the 20 bins and 19 in-transit exposures following the same strategy as used in \cite{h15}.  The weights are obtained by injecting the templates into the data at the start of the analysis at a low level as to not affect the statistics of the CCF \cite{brogi16}, and subsequently measuring the resulting increase of the CCF. Wavelength regions with poorer data quality or fewer spectral lines of the target species are thereby implicitly given a low weight.

The opacities of Fe, Fe$^+$, Ti, and Ti$^+$ \cite{kurucz17} are computed using standard methods \cite{rothman98,heng17} and the open-source \texttt{HELIOS-K} opacity calculator \cite{gh15}.  Equilibrium-chemistry calculations are performed using the \texttt{FastChem} computer code \cite{stock18}.  We note that the \texttt{JANAF} database (\texttt{https://janaf.nist.gov}) supplies the Gibbs free energies needed to perform the equilibrium-chemistry calculations up to only 6000 K.  Chemical kinetics calculations are performed using the \texttt{ARGO} computer code \cite{rh16}. The templates for the neutrals (Fe, Ti) and ions (Fe$^+$, Ti$^+$) assume temperatures of 3500 K and 4500 K, respectively; assuming different temperatures result in slight changes to the signal-to-noise of each detection.  The bright CCF streaks associated with each species are co-added along each streak, in the rest frame of the exoplanet, by using the measured orbital parameters \cite{gaudi17} and experimenting with a range of values for the orbital velocity of the exoplanet.  We estimate the SNRs of the individually detected species by measuring the standard deviation of the CCF at velocities more than 50 km s$^{-1}$ away from the exoplanet's rest frame.  We note that water was not detected below the 200 ppm level ($3 \sigma$).

\vspace{0.2in}

\noindent
{\scriptsize \textbf{Data availability:} The observations that support the findings of this study have been obtained as part of DDT program A35DDT4 (PI: Ehrenreich) and will become available in the public archive of the TNG after the data proprietary period of the TNG expires.

\vspace{0.2in}

\noindent
\textbf{Code availability:} All codes, programs and algorithms that were used to support the findings of this study are described in previously published literature. The codes to calculate the opacity functions and to model atmospheric chemistry are publicly available on \texttt{https://github.com/exoclime}.}

\vspace{0.2in}

\noindent
{\scriptsize \textbf{Acknowledgments:} This project has received funding from the European Research Council (ERC) under the European Union's Horizon 2020 research and innovation programme (projects {\sc Four Aces} and {\sc EXOKLEIN} with grant agreement numbers 724427 and 771620, respectively).  This work has been carried out in the framework of the PlanetS National Centre of Competence in Research (NCCR) supported by the Swiss National Science Foundation (SNSF).}

\begin{figure*}
\begin{center}
\includegraphics[width=1.8\columnwidth]{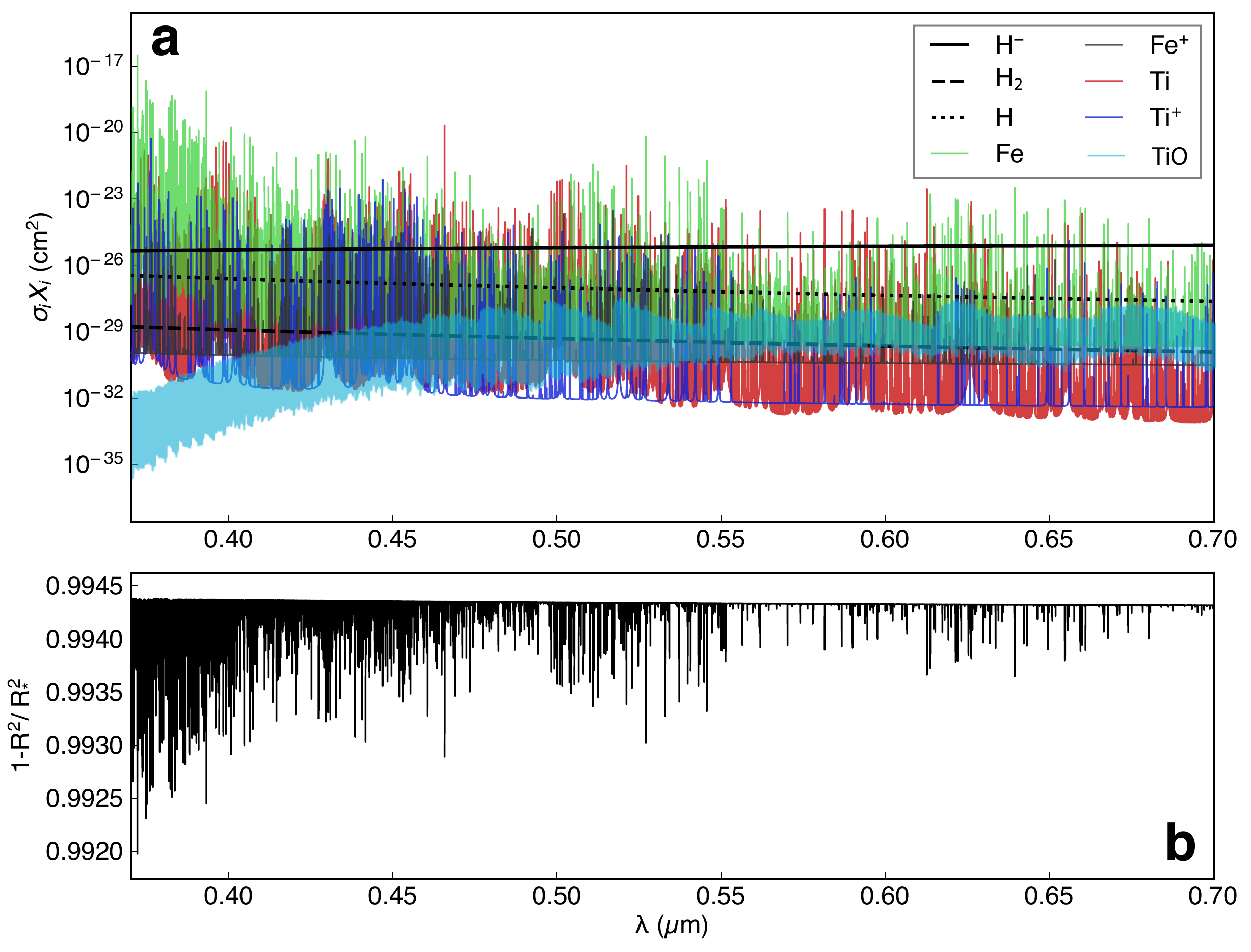}
\includegraphics[width=\columnwidth]{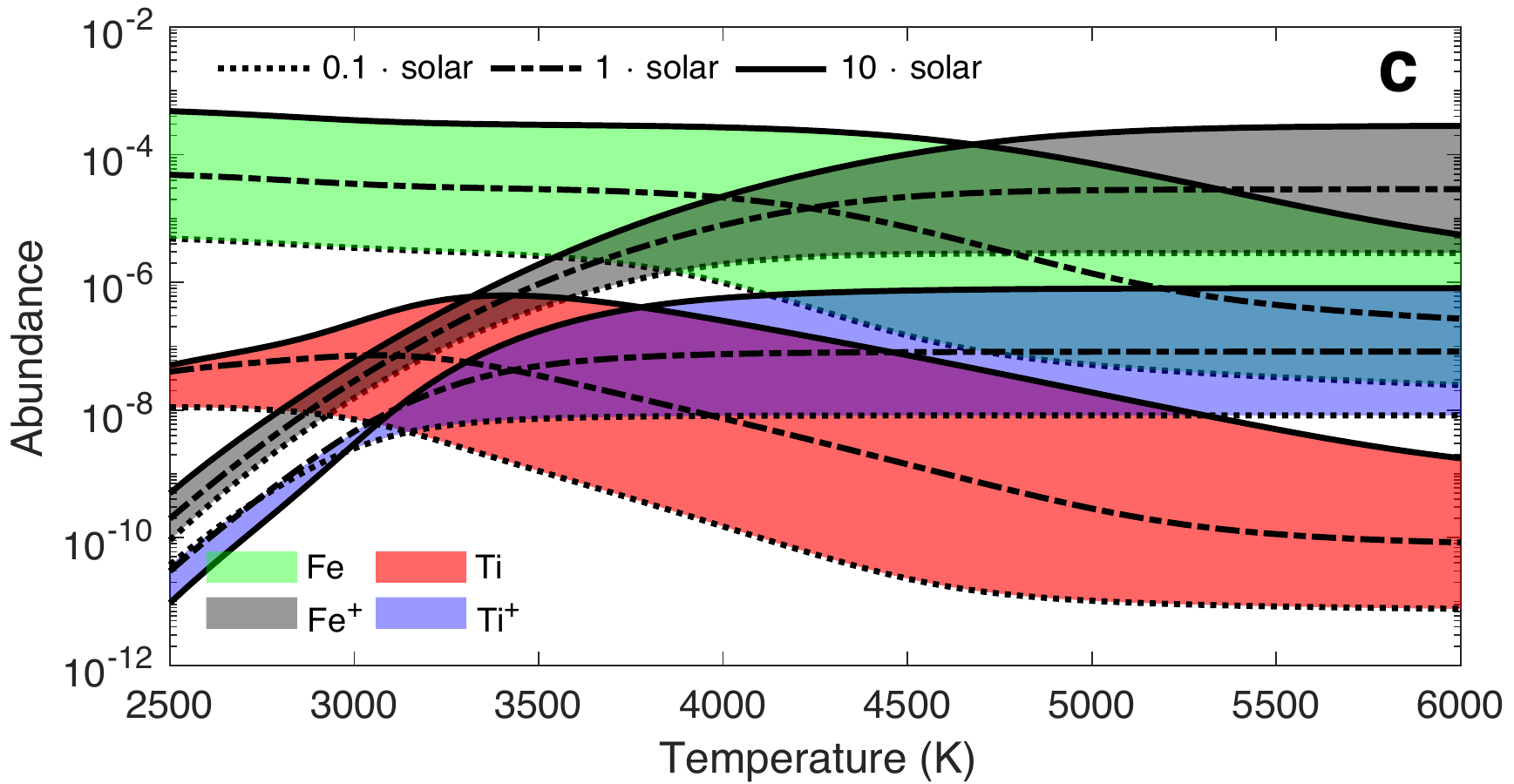}
\includegraphics[width=\columnwidth]{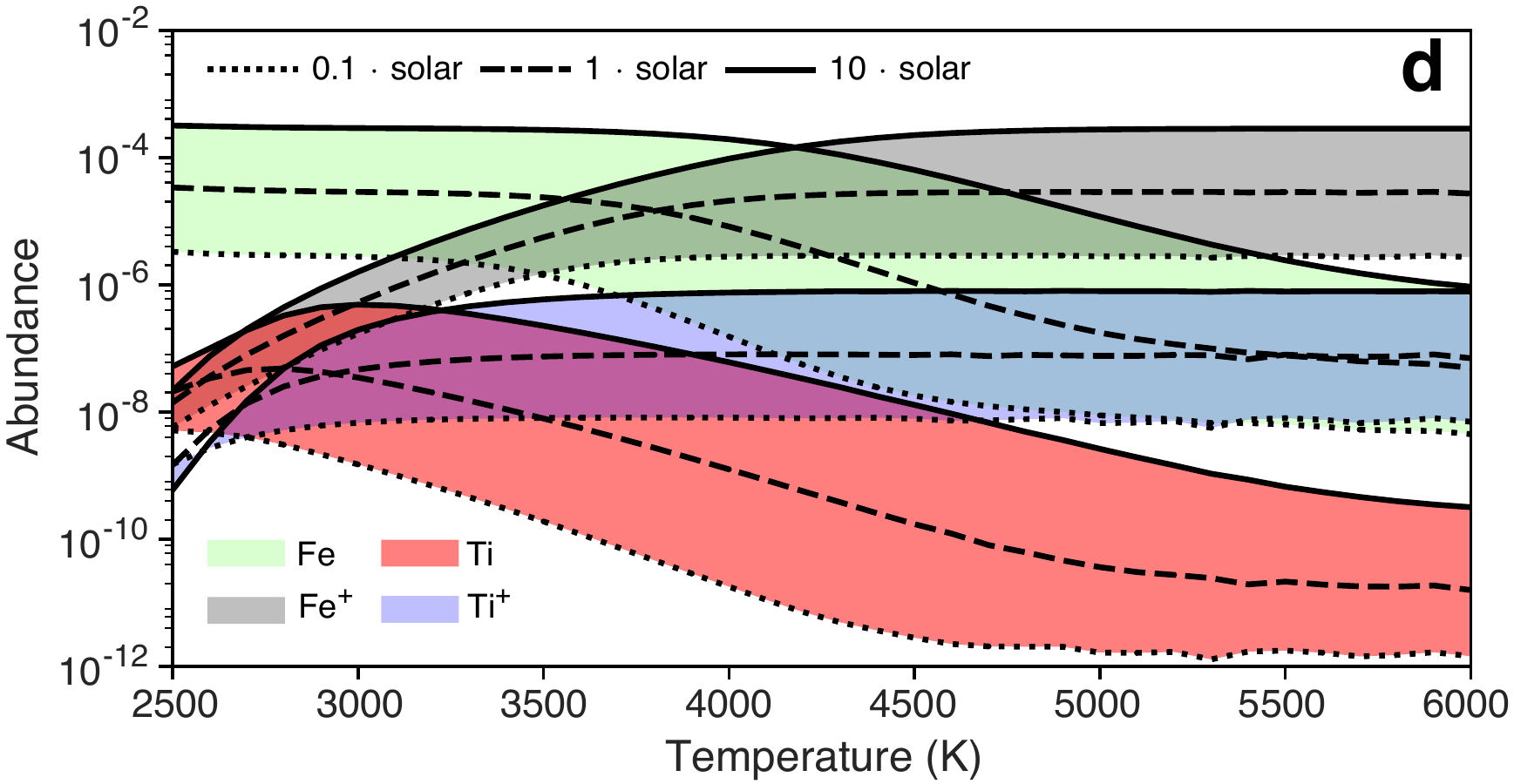}
\end{center}
\caption{\textbf{a:} Cross sections of Rayleigh scattering associated with atomic and molecular hydrogen, bound-free absorption associated with the hydrogen anion and the spectral lines of neutral and singly-ionized iron and titanium, as well as that of titanium oxide, weighted by their respective relative abundances by number (volume mixing ratios), assuming chemical equilibrium, a temperature of 4000 K and solar metallicity.  The dominance of H$^-$ absorption implies that one may use it to estimate the pressure associated with the transit chord probed in KELT-9b, which we compute to be about 10 mbar.  \textbf{b:} The theoretical transmission spectrum corresponding to the cross sections shown.  \textbf{c:} Mixing ratios of neutral and singly-ionized iron and titanium as functions of temperature, assuming chemical equilibrium and for metallicities from $0.1\times$ to $10\times$ solar.  \textbf{d:} Same as \textbf{c}, but including photochemistry and a representative vertical mixing strength of $10^{10}$ cm$^2$ s$^{-1}$.}
\label{fig:theory}
\end{figure*}

\begin{figure*}
\begin{center}
\includegraphics[width=1.8 \columnwidth]{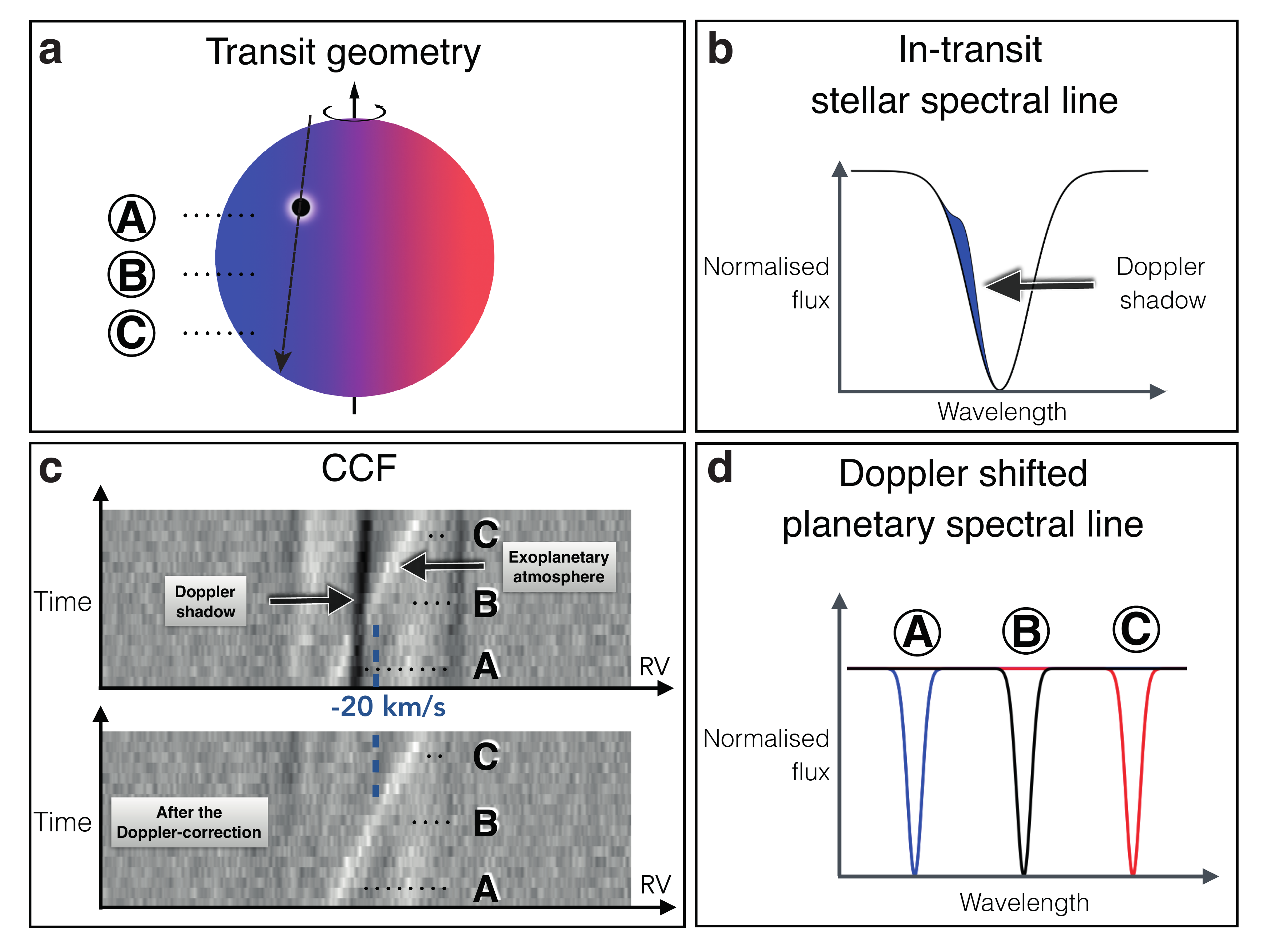}
\end{center}
\caption{Schematic of orbital geometry of exoplanet as it moves across the stellar disk (\textbf{a}).  The obscuration of part of the stellar disk shows up as an enhancement in flux of the stellar absorption line (\textbf{b}), which translates into a deficit in the cross correlation function (CCF).  Furthermore, as the exoplanet progresses in its orbit, its projected orbital velocity shifts from being blueshifted (at point A) to being redshifted (at point C) (\textbf{d}).  These two distinct signatures show up in the data (\textbf{c}) as a Doppler shadow (deficit in the CCF) and a bright streak (enhancement in the CCF).}
\label{fig:obs}
\end{figure*}

\begin{figure*}
\begin{center}
\includegraphics[width=2.0\columnwidth]{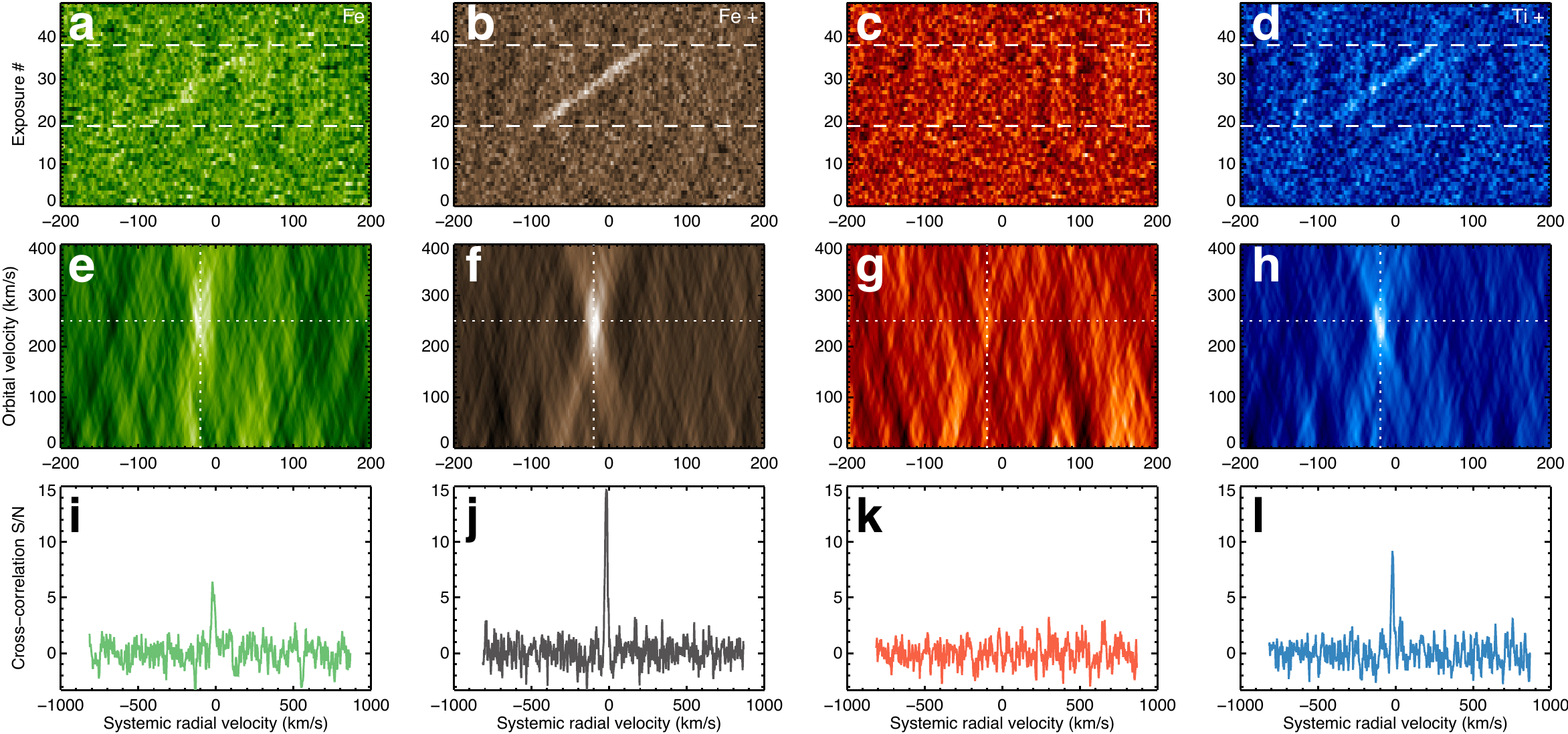}
\end{center}
\caption{Separate cross-correlation analyses using templates with neutral iron (Fe), singly-ionized iron (Fe$^+$), neutral titanium (Ti) and singly-ionized titanium (Ti$^+$).  Each template assumes a spectral continuum associated with the hydrogen anion (H$^-$).  \textbf{a--d:} The CCFs with the Doppler shadow subtracted.  The absorption spectrum of the exoplanetary atmosphere shows up as bright streaks.  The horizontal dashed lines mark the start and end of transit.  \textbf{e--h:} The data co-added, along each streak, in the rest frame of the exoplanet.  \textbf{i--l:} The cross correlation signal-to-noise ratio (SNR) extracted at the orbital velocity of 254 km s$^{-1}$.  Fe, Fe$^+$ and Ti$^+$ are clearly detected, but Ti is not.  The horizontal and vertical dotted lines denote the expected orbital velocity of Kelt-9b and the systemic velocity of the Kelt-9 system, respectively.}
\label{fig:detections}
\end{figure*}

\noindent
{\scriptsize HJH: led the data reduction and analysis, designed the cross-correlation templates, co-wrote the manuscript, co-led the scientific vision and coordination, sourced database information for opacity calculations and made all of the plots in Figure \ref{fig:detections}

\noindent
DE: principal investigator of the HARPS-N dataset (A35DDT4, PI: Ehrenreich), co-wrote the manuscript

\noindent
KH: led the scientific vision, coordination and interpretation, led writing and typesetting of the manuscript

\noindent
DK: performed chemical-equilibrium calculations using \texttt{FastChem} computer code and made the corresponding plot in Figure \ref{fig:theory}

\noindent
SLG: sourced database information for and performed opacity calculations using \texttt{HELIOS-K} computer code

\noindent
RA: performed the telluric subtraction using \texttt{Molecfit} \cite{smette15}, confirmed detections of Fe, Fe$^{+}$ and Ti$^{+}$ by repeating the cross-correlation analysis with binary masks

\noindent
RD: compiled cross sections for spectral lines and continuum and made the corresponding plot in Figure \ref{fig:theory}

\noindent
AW: led the data acquisition and observation, performed the extraction of the spectra, confirmed Doppler-shadow measurement and performed its removal

\noindent
MO: made schematic plot in Figure \ref{fig:obs}

\noindent
LP: co-designed the cross-correlation templates and performed additional checks on outcome

\noindent
PBR: performed chemical kinetics calculations using \texttt{ARGO} computer code and made the corresponding plot in Figure \ref{fig:theory}

\noindent
EM: granted a night out of his Director's Discretionary Time to make the project possible

\noindent
LDF: performed the TNG/HARPS-N observations}

\end{document}